\documentclass[preprintnumbers,amsmath,amssymb]{revtex4}

\usepackage{dcolumn}
\usepackage{bm}
\usepackage[dvipdfm]{graphicx}
\usepackage{color}

\begin{document}

\preprint{APS/123-QED}

\title{Renormalization group calculations for wetting transitions of infinite order and continuously varying order. I. Local interface Hamiltonian approach.}


\author{J.O. Indekeu$^1$, K. Koga$^2$, H. Hooyberghs$^1$ and A.O. Parry$^3$}

\affiliation{$^1$Institute for Theoretical Physics, KU Leuven,
BE-3001 Leuven, Belgium\\$^2$Department of Chemistry, Faculty of Science, Okayama University, Okayama 700-8530, Japan \\$^3$Department of Mathematics, Imperial College London, London SW7 2BZ, UK}.\\

\date{\today}

\begin{abstract}\noindent
We study the effect of thermal fluctuations on the wetting phase transitions of infinite order and of continuously varying order, recently discovered within a mean-field density-functional model for three-phase equilibria in systems with short-range forces and a two-component order parameter. Using linear functional renormalization group (RG) calculations within a local interface Hamiltonian approach, we show that the infinite-order transitions are robust. The exponential singularity (implying $2-\alpha_s = \infty$) of the surface free energy excess at infinite-order wetting as well as the precise algebraic divergence (with $\beta_s = -1$) of the wetting layer thickness are not modified as long as $\omega < 2$, with $\omega$ the dimensionless wetting parameter that measures the strength of thermal fluctuations. The interface width diverges algebraically and universally (with $\nu_{\perp} = 1/2$). In contrast, the non-universal critical wetting transitions of finite but continuously varying order are modified when thermal fluctuations are taken into account, in line with predictions from earlier calculations on similar models displaying weak, intermediate and strong fluctuation regimes. 
\end{abstract}

\maketitle

\section{Introduction}
In recent work \cite{Letter,Long} wetting transitions of infinite order were uncovered in a mean-field density functional theory (DFT) for systems with short-range forces and with a two-component order parameter. Although infinite-order wetting transitions were known to show up in certain fluctuation regimes studied using functional renormalization group (RG) theory for wetting or related methods \cite{FH,LN,BHL,LKZ,KL2,KL1}, it was a surprise that they can appear prominently already at mean-field level. A revisitation of a variety of early DFT's for wetting \cite{H,W} has led to the conclusion \cite{Long} that segments of infinite-order wetting transitions must be fairly ubiquitous, but have apparently long been overlooked, in models with a multi-component order parameter. These segments typically connect a regime of first-order wetting to one of critical wetting with continuously varying (non-universal) critical exponents.

We start with recalling briefly the main ingredients of the model, which has been described and discussed in detail in two earlier papers \cite{Letter,Long}. The mean-field DFT is defined through a functional $\hat\sigma$ of two spatially varying densities, or density components, $\rho_1({\bf r})$ and $\rho_2({\bf r})$. It represents the excess free energy per unit area of an interface, oriented perpendicular to $z$,
\begin{equation}\label{KWfunctional}
\hat\sigma[\rho_1,\rho_2] = \int_{-\infty}^{\infty}dz \left < \left \{
\frac{1}{2} (\nabla \rho_1({\bf r}))^2 +  \frac{1}{2} (\nabla
\rho_2({\bf r}))^2 + F(\rho_1({\bf r}),\rho_2({\bf r});a,b)\right \} \right >_{\{x,y\}}
,
\end{equation}
 The outer brackets denote that the integrand is averaged over the directions $x$ and $y$ parallel to the interface. The free-energy per unit volume $F$ is the following 6-th order polynomial
\begin{equation}\label{KWfunction}
 F(\rho_1,\rho_2;a,b) = \left ( (\rho_1 + 1)^2+\rho_2^2\right )\left(  ( \rho_1/a  )^2 + (\rho_2 -b)^2\right ) \left( (\rho_1-1)^2 + \rho_2^2\right ).
\end{equation}

The model parameter $a$ is an {\em asymmetry} variable, $a=1$ being its symmetric value. This presence of $a$ may have various grounds. In systems with an obvious geometrical symmetry relating the two densities $\rho_1$ and $\rho_2$, $a$ may be related to spatial anisotropy. This is the case, e.g., of a ferromagnet with cubic anisotropy \cite{W} or in general of systems that can be mapped onto a magnetic model with a magnetization vector order parameter \cite{H}. On the other hand, if the two densities are unrelated by any symmetry, it is always possible to redefine and scale them so that the {\em gradient-squared part of the functional is symmetric} (and diagonal) in $\rho_1$ and $\rho_2$, as it is in \eqref{KWfunctional}. Without loss of generality, our asymmetry parameter $a$ is defined adopting this convention. Different conventions may lead to different definitions for $a$ and consequently to differences in $a$-dependent calculational results, but do not affect the ultimate physical results (for the critical exponents, etc.). A model for which in this regard different conventions are used in different works, is, e.g., the Ginzburg-Landau DFT for superconductivity \cite{IvLLetter,IvL,Long,vLH}, in which one order parameter pertains to the superconducting wave function and the other to the magnetic vector potential. The model parameter $b$ is a control variable that allows wetting to be induced. Although it has the same dimension as the density $\rho_2$, it is physically a field-like variable which may depend on temperature and/or other external fields.

For arbitrary $a$ and $b$, $F$ reaches its minimum value, $F=0$, when the densities take their bulk-phase values at $z = \pm \infty$. These are
\begin{eqnarray}\label{bulkdensities}
\alpha \;\; \mbox{phase}&:& \rho_1 = -1,\; \rho_2 = 0; \nonumber \\
\beta \;\; \mbox{phase}&:& \rho_1 = 0,\;\;\;  \rho_2 = b; \\
\gamma \;\; \mbox{phase}&:& \rho_1 = 1,\;\;\;  \rho_2 = 0. \nonumber \\ \nonumber
\end{eqnarray}

As in the pertinent foregoing work \cite{Letter,Long} we study the wetting or nonwetting of the $\alpha\gamma$ interface by the $\beta$ phase. In the DFT defined through (\ref{KWfunctional}) the wetting transition was found to be of first order for $a>1$. A second-order wetting transition was found for $a=1$ (symmetric model), when $b$ is lowered towards $b_w(1) \equiv 0.681...$, already in \cite{KogaWidom2008}. For that transition the critical exponent associated with the free-energy singularity, $2-\alpha_s$, takes the value 2, and the critical exponent of the wetting layer thickness takes the value $\beta_s = 0($log$)$, signifying a logarithmic divergence at wetting. Further, a non-universal critical wetting transition was obtained for $a < 1$, upon lowering $b$ towards $b_w = 0$. For this transition, $2-\alpha_s = 1/(1-a)$ and $\beta_s = 0($log$)$. Finally, a segment of infinite-order wetting transitions was found at $a=1$ and for $0 < b < b_w(1) = 0.681...$. The singularity in the spreading coefficient near wetting was conjectured to be of the form, in the limit $a\uparrow 1$,
\begin{equation}\label{SInfExp}
-S \propto e^{-C/(1-a)}; \;\; \mbox{with} \;\; C>0,
\end{equation}
implying $2-\alpha_s = \infty$, and the wetting layer thickness was conjectured to diverge in the {\em algebraic} manner,
for $a\uparrow 1$,
\begin{equation}\label{ellalg}
\ell  \propto (1-a)^{-1},
\end{equation}
implying $\beta_s = -1$.
These conjectures were based on the analytical solution of a different but related model, for which the leading terms, for large $\ell$, of an interface potential $V(\ell)$ could be calculated. Furthermore, accurate numerical computations for the original model support the validity of the results \eqref{SInfExp} and \eqref{ellalg}. The global wetting phase diagram in the $(a,b)$-plane has been presented in \cite{Letter,Long} and is, for the sake of clarity, reproduced in Fig.1 in annotated version.
\begin{figure}
\includegraphics[scale=1]{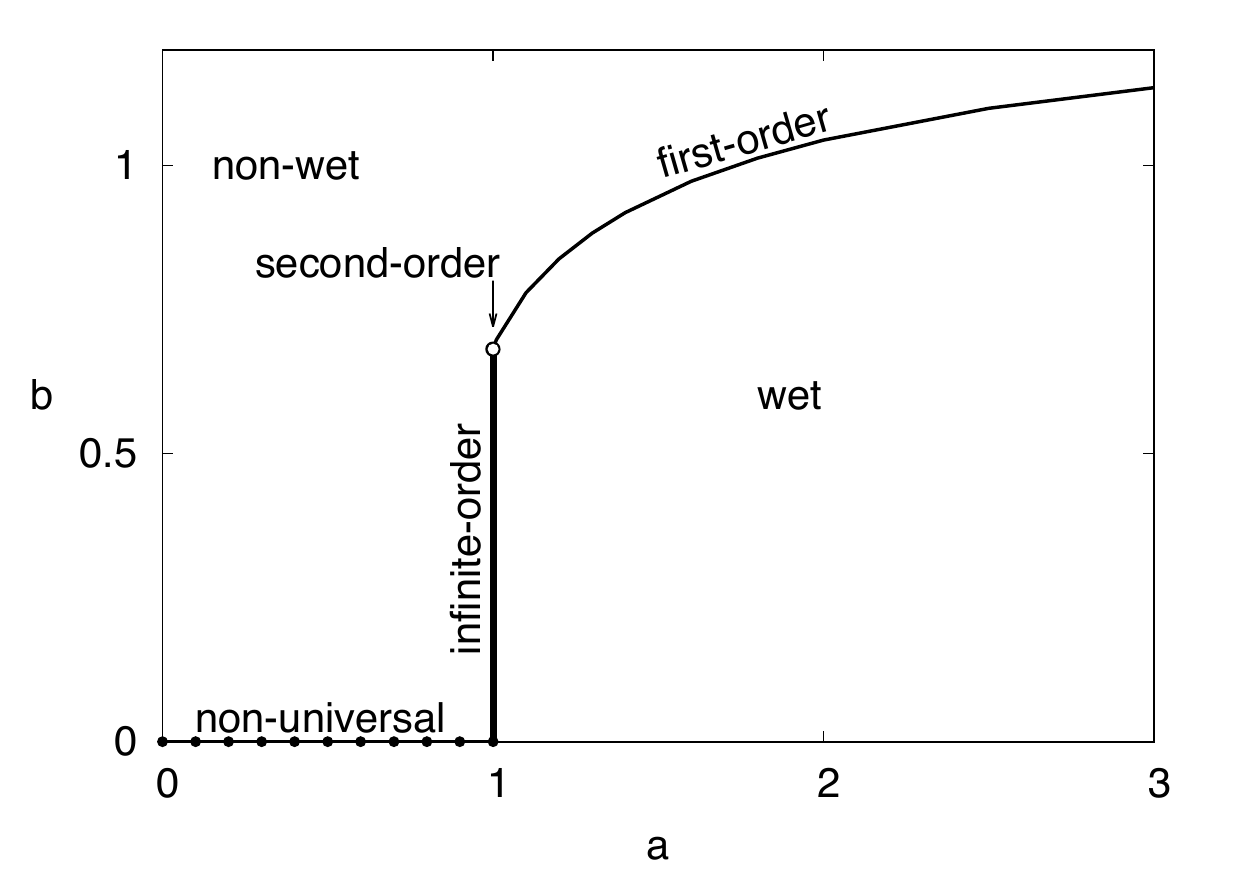}
\caption{The global wetting phase diagram of the model in the $(a,b)$-plane, reproduced from \cite{Letter,Long}. Indicated are the line of first-order wetting ($a>1$), the second-order wetting point ($a=1$ and $b= 0.681...$), the line of infinite-order wetting ($a=1$) and the line of wetting transitions of continuously varying order (``non-universal", $b=0$). These wetting phase boundaries separate non-wet from wet equilibrium states.}
\end{figure}

The remainder of this paper is organized as follows. In Section II we discuss the occurrence of infinite-order wetting transitions in mean-field theories and RG theories. Section III presents a new derivation of an interface potential $V(\ell)$ using the simplest possible crossing criterion for the interface position. Section IV is devoted to RG calculations for wetting transitions that are of infinite order already at mean-field level. In Section V we point out that the renormalization of the wetting transitions of finite but continuously varying order qualitatively reproduces previously obtained RG corrections for similar wetting transitions with, however, interesting quantitative differences. Conclusions are drawn in Section VI. 

\section{Occurrences of infinite-order wetting}
\subsection{Mean-field theories}
One of the main conclusions of \cite{Long} is that every mean-field DFT whose properties in the vicinity of the $\beta$ phase can be ``mapped" onto those of \eqref{KWfunctional} is susceptible of displaying a segment of infinite-order wetting along the ``symmetric" line $a=1$, spanning some range of field variables. However, it must be checked case by case whether the line $a=1$ belongs to the physical subspace in which wetting transitions are possible. Let us comment in this regard on the three examples discussed in \cite{Long}.

In the pioneering two-component order parameter DFT of Hauge \cite{H} for wetting at a wall the asymmetry parameter is the ratio of two curvatures, $\lambda_1$ and $\lambda_2$, that characterize the shape of the potential $-F$ in the vicinity of the bulk wetting-phase point ${\bf M} = (-M,0)$ in the plane of the order parameters ${\bf M} = (M_1,M_2)$. For $a \equiv \lambda_1/\lambda_2 >1$ the wetting transition is of first-order. For $a=1$ it is a standard second-order transition ($\alpha_s=0$), for $1/2 < a < 1$ it is of higher order and non-universal ($2-\alpha_s = 1/(1-a)$), and for $0<a<1/2$  is (again) a standard second-order transition ($\alpha_s=0$). However, recent insights \cite{Long} imply that precisely at $a=1$ the regimes of first- and higher-order wetting are connected by a segment of infinite-order transitions. We infer that this segment extends, in the notation of \cite{H}, from $\tau = 0$ to $\tau = -H_2\alpha_2/\alpha_1 (>0)$, where $\tau \equiv cM + H_1$ is the control parameter that allows one to induce wetting by varying the temperature, through $M$, and/or by varying the surface field ${\bf H} = (H_1,H_2)$. Here, $\alpha_2/\alpha_1^{\lambda_2/\lambda_1} = a_{0,2}/a_{0,1}^{\lambda_2/\lambda_1} (>0)$, with $a_{0,1}$ and $a_{0,2}$ parameters associated with the ``free" interface between the wetting and non-wetting phases in bulk and $c(>0)$  a surface enhancement parameter. We note that in this model the interface potential features the following leading two terms, for $\lambda_1 \leq \lambda_2$ (critical wetting)
\begin{equation}\label{Vlambda}
V(\ell) = - \tau \alpha_1 e^{-\lambda_1 \ell} - H_2 \alpha_2 e^{-\lambda_2 \ell} + {\cal O}(e^{-2\lambda_1 \ell}),
\end{equation}
with $\ell$ a suitably scaled wetting layer thickness.

In the two-component order parameter DFT of Walden {\em et al.} \cite{W} for investigating the surface properties of a ferromagnet with cubic anisotropy and vector order parameter ${\bf M} = (M_x,M_y)$, a similar scenario unwinds. The asymmetry parameter $a$ is again the ratio of two curvatures, that characterize the shape of the potential $-F$ in the vicinity of the bulk wetting-phase point ${\bf M}_A = (\sqrt{-t},0)$ in the plane of the order parameter components. Here, $t$ is the usual reduced temperature distance to the Curie point, $(T-T_c)/T_c$. This ratio $a$ equals the ratio of two lengths, $a \equiv  \xi_y/\xi_x = \sqrt{2/(\lambda-1)}$, with $\lambda$ a measure of the magnetic anisotropy ($\lambda = 1$ being the {\em isotropic} value). Note that the {\em symmetric} value for $a$, $a=1$, corresponds, however, to $\lambda =3$, which is already well within the range of values of $\lambda$ of materials with ``cubic anisotropy", $\lambda > 1$, to which our attention is restricted from now on. The anisotropy is (strongly) temperature dependent \cite{Asselin}. For $a >1 $ $(1 < \lambda < 3)$ the wetting transition is of first order. For $a=1$ $ (\lambda=3)$ it is of second order ($\alpha_s=0$), for $1/2 < a < 1 $ $(3 < \lambda < 9)$ it is of higher order and non-universal ($2-\alpha_s = 1/(1-a)$), and for $0<a<1/2 $ $(\lambda > 9) $ it is universal and of second order ($\alpha_s=0$). Also for this model, the revisitation in \cite{Long} concluded that precisely at $a=1$ the regimes of first- and higher-order wetting must be connected by a segment of infinite-order transitions. We infer that this segment extends, in the notation of \cite{W}, from $\tau = 0$ to $\tau = H_y^s a_{0y}/a_{0x} (>0)$, where $\tau \equiv \xi_0 c M_{A,x} + H_x^s$ is the control parameter that allows one to induce wetting by varying the temperature, through $t$, and/or by varying the surface field ${\bf H^s} = (H_x^s,H_y^s)$. Here, $a_{0y}/a_{0x}(>0)$ is a ratio of parameters associated with the ``free" interface between the wetting and non-wetting phases in bulk, $c(>0)$ is a surface enhancement parameter and $\xi_0 $ is a (constant) length. In this model the interface potential features the following leading two terms, for $\lambda \geq 3$ (critical wetting)
\begin{equation}\label{Vxi}
V(\ell) = - \tau \xi_0 a_{0x} e^{- \ell/\xi_x} + \xi_0 H_y^s a_{0y} e^{- \ell/\xi_y} + {\cal O}(e^{-2 \ell/\xi_x}).
\end{equation}

In the Ginzburg-Landau (GL) theory for superconductivity \cite{IvLLetter,IvL,vLH,CI}, after suitable scaling of the two order parameters so as to arrive at a symmetric gradient-squared part of the functional, the asymmetry parameter takes the form $ a = \kappa \sqrt{2}$, with $\kappa = \lambda/\xi$ the GL parameter, being the ratio of the superconducting coherence length to the magnetic penetration depth. Close to the bulk critical point $\kappa$ is a material constant, with $0 < \kappa < 1/\sqrt{2}$ applicable to type-I superconductors, for which wetting by a macroscopic superconducting layer (Meissner phase), intruding between the normal phase and the sample surface, can be investigated. For materials with $0 < \kappa < 0.374$ first-order wetting is possible \cite{IvLLetter,IvL}, while for $0.374 < \kappa < 1/\sqrt{2}$ non-universal critical wetting is possible \cite{IvLLetter,IvL,vLH} with $2-\alpha_s = 1/(1-\kappa \sqrt{2})$.
At first sight it would seem that the global wetting phase diagram for this system might feature a segment of infinite-order transitions at $a=1 $ $ (\kappa = 1/\sqrt{2})$, but this is {\em not} the case. In the limit $\kappa \uparrow 1/\sqrt{2}$ the surface tension between the superconducting and normal phases vanishes (bulk multicritical point) and the notion of wetting phase transition ceases to exist. Nevertheless, phenomena reminiscent of enhanced adsorption do occur in this limit \cite{CI,Dietrich,Speth}. This DFT thus provides an interesting exception to the scenario of infinite-order wetting in DFT's with two order parameters. For completeness, we also give the leading structure of the interface potential, for $\kappa < 1/\sqrt{2}$,
\begin{equation}\label{Vkappa}
V(\ell) = - A e^{- \kappa \sqrt{2} \,\ell} + B e^{- \ell} + {\cal O}(e^{-2 \kappa \sqrt{2} \,\ell}),
\end{equation}
with $A$ and $B$ given explicitly in \cite{vLH}.

The occurrence of infinite-order wetting in DFT's with continuously varying critical exponents compellingly raises the question whether infinite-order wetting occurs also in the famous van der Waals theory of wetting for short-range forces with a {\em one-component} order parameter \cite{AH}, featuring a {\em non-universal} $2-\alpha_s$ that depends on the ratio of the inverse bulk correlation length $\lambda$ to the inverse range $\beta$ of the exponentially decaying wall-fluid potential.
In this model the range of the exponentially decaying effective fluid-fluid potential is set to unity, and the bulk correlation length varies monotonically from unity at $T=0$ to infinity at the bulk critical point at $T=T_c$. Consequently, at finite temperatures, $\lambda < 1$. For $\beta = 1$, the second-order wetting transition of the Sullivan model is recovered with $2-\alpha_s = 2$ and $\beta_s = 0(log)$. For $\beta \neq 1$ there are several possibilities. Consider first the case $\beta < 2\lambda$. For $\beta < 1$ the wetting transition, {\em if} it exists, turns out to be of first order, while for $1 < \beta < 2 \lambda $ it is critical and non-universal with $2-\alpha_s = 1/(1-\lambda/\beta)$. The order of the transition may thus be high but {\em cannot diverge} since $\beta > 1$ and $\lambda \leq 1$. Next, for $\beta \geq 2\lambda$ (for arbitrary $\beta > 0$) the transition, {\em if} it is critical, settles on the usual universal second-order transition. This model therefore constitutes another ``counterexample" for which the would-be infinite-order transition (mathematically conceivable at $\lambda/\beta = 1$) does not take place since its location would fall outside the domain of validity of continuous wetting \cite{nonthermo}. Note that the border point of this domain lies at $\beta = 1$, for which only second-order wetting is possible, regardless of the value of $\lambda (\leq 1)$. We close this case by recalling that the interface potential for this model takes the form, for large $\ell$,
\begin{equation}\label{Vell}
V(\ell) = K_{\lambda}e^{- \lambda \ell} + K_{\beta} e^{- \beta \ell} + {\cal O}(e^{-2 \lambda \ell}),
\end{equation}
with analytic expressions for $K_{\lambda}$ and $K_{\beta}$ given in \cite{AH}.
\subsection{Renormalization Group theories and exact results}
The first infinite-order wetting transition that enjoyed some attention was that associated with the strong thermal fluctuation regime, $\omega > 2$, of the short-range critical wetting (SRCW) transition in {\em three} dimensions \cite{LKZ,BHL,FH}. Here, $\omega$ is the ``wetting parameter" given by
\begin{equation}\label{omega}
\omega = \frac{kT}{4\pi\sigma \xi^2},
\end{equation}
with $\sigma$ the interfacial tension between the wetting phase and the bulk phase far from the ``wall", and $\xi$ the bulk correlation length in the wetting phase. For the Ising model near $T_c$, the universal value $\omega \approx 0.8$ applies \cite{PREL}.
The pioneering functional RG treatments or related variational approaches of SRCW in $d=3$ \cite{BHL,LKZ} already revealed that the transition turns infinite order for $\omega > 2$, and, complemented with subsequent refined calculations \cite{FH}, provided the famous non-universal critical exponents in the weak ($0<\omega < 1/2$) and intermediate ($1/2<\omega<2$) fluctuation regimes relevant to the Ising model universality class. Note that the results of the mean-field approximation are formally retrieved in the limit $\omega \downarrow 0$. Interestingly, the global phase diagram of SRCW in $d=3$ (see Fig.2 in \cite{FH}) bears some resemblance to that of the mean-field DFT we discuss, in that a regime of transitions of continuously varying order culminates in an infinite-order transition when $2-\alpha_s$ diverges, for $a \uparrow 1$ in the mean-field DFT and for $\omega \uparrow 2$ in the functional RG theory. The algebraic divergence of the wetting layer thickness at wetting (with $\beta_s = -1$) is also shared by the MF model and the RG theory. However, the similarity is not complete. In the mean-field DFT the segment of infinite-order transitions is associated with a jump, at $a=1$, in the control parameter, say temperature and/or surface field, with which wetting can be induced, while in the RG treatment the control parameter at wetting, say $T_w$, varies continuously (linearly) as a function of $\omega$ for $\omega >2$.

In {\em two} dimensions an interesting segment of infinite-order wetting  transitions has been uncovered in the so-called intermediate fluctuation regime \cite{LN}, by means of Feynman path integral and transfer matrix methods, following pioneering analytical work which already established an infinite-order transition \cite{KL2}. In $d=2$ the fluctuation-induced repulsion between an unbinding interface and a wall, is, unlike in $d=3$, not exponentially but algebraically decaying, in the manner $V_{\rm fluc}(\ell) \propto 1/\ell^2$. When this entropic repulsion competes with an attractive direct interaction behaving for large $\ell$ as $-w/\ell^2$, with $w>0$, which is appropriate for certain systems with long-range forces, a wetting transition of infinite order is possible, with $2-\alpha_s = \infty$ and $\beta_s = -\infty$ (essential singularities for the spreading coefficient as well as for the wetting layer thickness). The global wetting phase diagram (see Fig.1 in \cite{LN}) displays a segment of infinite-order wetting at $w=1/4$, spanning an (infinite) jump in the amplitude of the short-range direct interaction between interface and wall.

\section{Derivation of an interface potential within mean-field theory}

Let us recall the starting point of the interface potential approach adopted in \cite{Long}. Expanding the surface free-energy functional \eqref{KWfunction} about the bulk $\beta$-phase point $(0,b)$ in the $(\rho_1,\rho_2)$-plane leads to the approximation
\begin{equation}
F(\rho_1,\rho_2;a,b) \approx F^{(2)}(\rho_1,\rho_2;a,b) \equiv (1+b^2)^2 \left[ (\rho_1/a)^2 + (\rho_2 - b)^2 \right].
\end{equation}
Solving the Euler-Lagrange equations \cite{Long} within this harmonic approximation leads to, with $Z \equiv A^{1/2} z$, where $A \equiv 2(1+b^2)^2$,
\begin{eqnarray}\label{first}
\rho_1(Z) &= &a_1 \exp(Z/a) + b_1 \exp(-Z/a), \nonumber \\
\rho_2(Z) &= &b +a_2 \exp(Z) + b_2 \exp(-Z),
\end{eqnarray}
and we define $b_0^{(2)} \equiv \rho_2(0)$. Clearly, symmetry considerations invite us to define $Z=0$ as the middle plane of the layer of $\beta$-phase material, so  $\rho_1(0) = 0$ and $\dot\rho_2(0) = 0$, from which follow $b_1 = -a_1$ and $b_2=a_2$. Instead of determining the remaining two free parameters by the criteria pursued in \cite{Long}, we add a crossing criterion at $Z_{\times} \equiv L/2$, with $L \equiv A^{1/2} \ell$,
\begin{eqnarray}
\rho_1(L/2) &= &\rho_{1,\times}, \nonumber \\
\rho_2(L/2) &= &qb,
\end{eqnarray}
assuming that the width of the $\beta$ layer, $L$, is sufficiently large for the second equation to possess a solution (i.e. $b_0^{(2)} > qb$). Note that $\rho_{1,\times} \in {[0,1]}$ and $q \in {[0,1]}$ are, for the time being, free parameters. This crossing criterion is quite different from the trajectory-intersection strategy proposed in \cite{Long}. In particular, the trajectory intersection in the model in \cite{Long} always occurs near the $\beta$-phase point, while the crossing occurs, as we shall see, roughly half a wetting layer thickness ``away" from that point. Further, the crossing criterion relates the remaining free parameters $a_1$ and $a_2$ to the parameters $\rho_{1,\times}$ and $q$, for an arbitrarily chosen value of $L$,  through the simple relations
\begin{eqnarray}
a_1 &= &\frac{\rho_{1,\times}}{2 \sinh( L/2a)}, \nonumber \\
a_2 &= &- \frac{b}{4 \cosh (L/2)}.
\end{eqnarray}
The constrained order parameter solutions then take the simple analytic forms:
\begin{eqnarray}\label{firstexplicit}
\rho_1(Z) &= & \rho_{1,\times} \frac{\sinh(Z/a)}{\sinh(L/2a)},  \nonumber \\
\rho_2(Z) &= &b - (1-q)b \frac{\cosh(Z)}{\cosh(L/2)}.
\end{eqnarray}

In the spirit of a ``double parabola approximation" we now proceed to apply the harmonic approximation also about the $\gamma$-phase point $(1,0)$ in the $(\rho_1, \rho_2)$-plane. This leads to
\begin{equation}
F(\rho_1,\rho_2;a,b) \approx F^{(2)}(\rho_1,\rho_2;a,b) \equiv 4 \left(\frac{1}{a^2}+b^2\right) \left[ (\rho_1-1)^2 + \rho_2 ^2 \right].
\end{equation}
The Euler-Lagrange equations are solved by, with $Z' \equiv C^{1/2} z'$, where $C \equiv 8(1/a^2+b^2)$,
 \begin{eqnarray}\label{second}
\rho_1(Z') &= &{1  - c_1 \exp(-Z')}, \nonumber \\
\rho_2(Z') &= & c_2 \exp(-Z'),
\end{eqnarray}
where we already implemented the boundary conditions appropriate to the $\gamma$ phase point, which must be reached for $Z' \rightarrow \infty$. 
The two approximate pairs of solutions, one valid near $\beta$ and the other valid near $\gamma$, can be matched at $Z'=0$ in the solutions \eqref{second}, which corresponds to $Z=L/2$ in the solutions \eqref{first}. This implies
\begin{eqnarray}
c_1&= &{1-\rho_{1,\times}}, \nonumber \\
c_2&=&qb. 
\end{eqnarray}
In this way, a requirement of continuity of the order parameters at the crossing point fixes the parameters $c_1$ and $c_2$ in terms of two remaining freedoms. While the order parameters are continuous, their derivatives need not be, at the matching point in the $(\rho_1, \rho_2)$-plane.

Within this double-parabola approximation we now define, as usual, the interface potential $V(L)$ as the constrained surface free energy of a layer of $\beta$ of finite thickness $L$ adsorbed at the $\alpha$$\gamma$ interface, minus the surface free energy of an infinitely thick layer (the wet profile). The first part of the surface free energy cost is obtained by evaluating the functional \eqref{KWfunctional} between the limits $z=0$ and $z=\ell/2$ in the solutions \eqref{first} and the second part is obtained by evaluating the functional \eqref{KWfunctional} between the limits $z'=0$ and $z'=\infty$ in the solutions \eqref{second}. Note that the second part is independent of $L$. We write, suggestively,
\begin{eqnarray}
V(L)/2 \equiv \sigma^{(2)}_{\beta,[0,L/2]}  + \sigma^{(2)}_{\gamma,[0,\infty]} - (\sigma^{(2)}_{\beta,[0,\infty]}  + \sigma^{(2)}_{\gamma,[0,\infty]}),
\end{eqnarray}
where $\sigma$ denotes the functional $\hat\sigma$ evaluated in the optimal profiles. Calculation entails
\begin{eqnarray}\label{intpot}
V(L) = \sqrt{A} \left ( -(1-q)^2 b^2 \frac{e^{-L}}{1 +  e^{-L}} + \frac{\rho_{1,\times}^2}{a} \frac{e^{-L/a}}{1 -  e^{-L/a}} \right ), 
\end{eqnarray}
which has a hard-wall divergence for $L \downarrow 0$. 
On the other hand, the surface free energy of the wet profile corresponds to (twice) the interfacial tension of the $\beta\gamma$ interface. In this harmonic approximation this reads
\begin{eqnarray}
\sigma^{(2)}_{\beta\gamma}  \equiv    \sigma^{(2)}_{\beta,[0,\infty]}  + \sigma^{(2)}_{\gamma,[0,\infty]},
\end{eqnarray}
and we obtain
\begin{eqnarray}
\sigma^{(2)}_{\beta\gamma}  = \frac{\sqrt{A}}{2} \left ((1-q)^2b^2  + \frac{\rho_{1,\times}^2}{a}\right )   + \frac{\sqrt{C}}{2} \left (q^2b^2  + (1-\rho_{1,\times})^2\right ). 
\end{eqnarray}

It is interesting to examine the precise form of the wet trajectory $\rho_2(\rho_1)$ in the $(\rho_1,\rho_2)$-plane. This trajectory consists of two parts. The first part, from the $\beta$ phase point till the crossing point, is most easily obtained by recasting the solutions \eqref{first} in a form which is suitable for starting from the $\beta$ phase point for $z' \rightarrow -\infty$ and reaching the crossing point at $z'=0$,
\begin{eqnarray}\label{firstagain}
\rho_1(Z') &= &\rho_{1,\times} \exp(Z'/a), \nonumber \\
\rho_2(Z') &= &b\left(1-(1-q)\exp(Z') \right),
\end{eqnarray}
which implies
\begin{eqnarray}
\rho_2 = b-(1-q) b\left (\frac{\rho_{1}}{\rho_{1,\times} } \right ) ^a,\;\;\mbox{for}\;\rho_{1} \leq \rho_{1,\times}.
\end{eqnarray}
The second part of the wet trajectory follows directly from \eqref{second} and takes the form
\begin{eqnarray}
\rho_2 = qb \frac{1- \rho_{1}}{1-\rho_{1,\times} },\;\;\mbox{for}\;\rho_{1} \geq \rho_{1,\times},
\end{eqnarray}
which is a straight line. It is interesting to note that the slope at $\rho_{1} = \rho_{1,\times}$ jumps from $-a(1-q)b/\rho_{1,\times}$ to $-qb/(1-\rho_{1,\times})$. Now requiring continuity of the slope of the wet trajectory at $\rho_{1,\times}$, allows us to eliminate one freedom. We thus obtain
\begin{equation} \label{opt}
q = \frac{a(1-\rho_{1,\times})}{\rho_{1,\times}+a(1-\rho_{1,\times})}.
\end{equation}
Note that the non-wet trajectory will, in general, display a discontinuity in slope at $\rho_{1,\times}$, whose magnitude depends also on $L$. 
A next opportunity to eliminate a free parameter is provided
by asking that the interfacial tension $\sigma^{(2)}_{\beta\gamma} $ be minimal with respect to $\rho_{1,\times}$, after substitution of \eqref{opt}. One readily checks that the minimum is reached for
\begin{equation}\label{minten}
\rho_{1,\times} = \left (1 + \frac{1+b^2}{2\sqrt{1+a^2b^2}}\right )^{-1},
\end{equation}
which takes the value 2/3 at $b=0$ (for any $a$) and 0.623 at an endpoint of the line of infinite order transitions, $a=1$ and $b = 0.681...$. In conclusion, adopting this criterion for fixing $\rho_{1,\times}$ would lead to a value that is rather insensitive to the parameters $a$ and $b$ in the region of our interest. We shall see further that the physical results we will derive are largely independent of the precise value of $\rho_{1,\times}$. 
Figure 1 illustrates our harmonic approximations supplemented with the crossing criterion and provides a comparison with the numerically exact trajectories, for both non-wet and wet states. 
\begin{figure}
\includegraphics[scale=1]{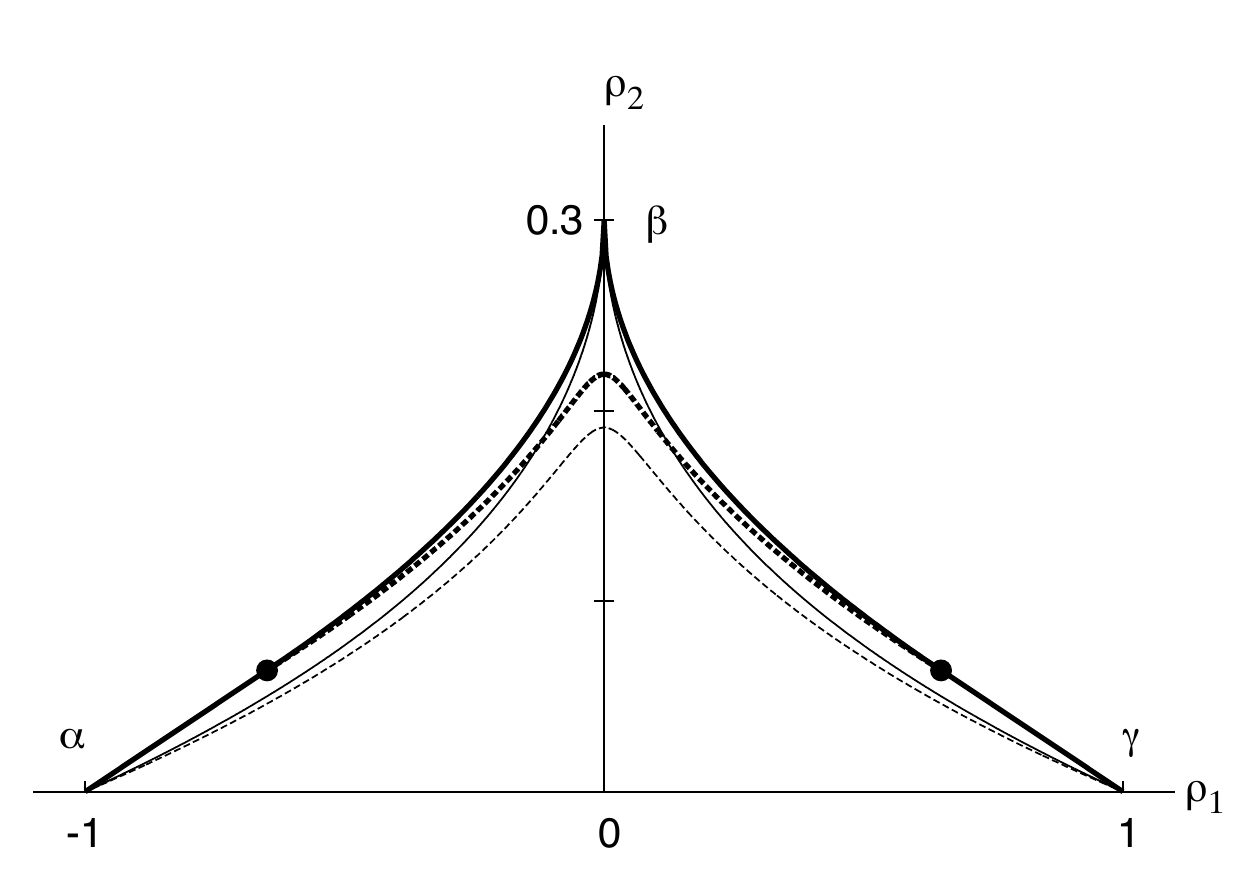}
\caption{The wet trajectory (thick solid line) and the non-wet trajectory (thick dashed line for
$|\rho_1| < \rho_{1,\times}$ and thick solid line for $ \rho_{1,\times} < |\rho_1| < 1$) within the harmonic approximation to the model and the corresponding numerically exact wet and non-wet trajectories (thin solid and dashed lines, respectively). The crossing points are marked by dots. For this illustrative case the following parameter values were used: $a=1/2$ and $b= 0.3$. The value of $L$ that minimizes the surface free energy of the non-wet trajectory within the harmonic approximation (thick dashed line) is given by $L = 3.4723...$. Note that the relation between $b$, $b_0^{(2)}$, and $L$, within the harmonic approximation, is given by $b - b_0^{(2)} = (1-q)b /\cosh(L/2)$.}
\end{figure}

From \eqref{intpot} we now derive the asymptotic form, for large $L$, of the interface potential $V(L)$, in the domain $0<a<1$ appropriate to critical wetting, 
\begin{eqnarray}\label{Vexpagen}
V(L)/\sqrt{A} =  - (1-q)^2b^2 \;e^{-L} + \frac{\rho_{1,\times}^2}{a}  \;e^{-L/a} + (1-q)^2b^2  \;e^{-2 L} +  {\cal O}(e^{-2L/a}, e^{-3L} ),
\end{eqnarray}
with $L$ the (scaled) thickness of the wetting layer of phase $\beta$ intruding between phases $\alpha$ and $\gamma$. We recall that $L$ is measured in units of the scaled distance $Z = z \sqrt{2(1+b^2)^2}$ and the factor of 2 absorbed in our definition of $L$ allows for the fact that the actual thickness of the wetting layer of $\beta$ corresponds, by symmetry, to about twice the ``wetting layer thickness" defined in the calculations of \cite{Long,note}.  Note that this interface potential differs qualitatively from the one engineered in \cite{Long}. Our present derivation is simpler and more transparent. While the difference is irrelevant at the level of the mean-field results due to the precise dependence on $b$ of the coefficient of the third term in $V(L)$ (i.e., the mean-field critical exponents are not modified), the difference can become important at the level of the RG calculations. 

Let us briefly derive the mean-field critical exponents from \eqref{Vexpagen}. Clearly, for $a>1$ we retrieve consistency with a first-order wetting transition, because the leading term then becomes positive. For $a \leq 1$ critical wetting is possible. For $a=1$ a universal second-order wetting transition is retrieved from the interface potential
\begin{eqnarray}\label{Vexpagenais1}
\frac{V(L)}{\sqrt{A}} =  \left(- (1-q)^2b^2  +\rho_{1,\times}^2 \right)\;e^{-L} +  \left((1-q)^2b^2  +\rho_{1,\times}^2\right) \;e^{-2 L} +  {\cal O}(e^{-3L} ),
\end{eqnarray}
which predicts that, in this harmonic approximation scheme, complete wetting takes places for $b$ less than the threshold
\begin{eqnarray}
b_w^{(2)}(1) =  \rho_{1,\times}/(1-q) = 1,
\end{eqnarray}
using \eqref{opt} at $a=1$, 
while the exact value is $b_w(1) = 0.681...$ \cite{KogaWidom2008}. Here we see explicitly that the value of the free parameter $\rho_{1,\times}$ in the crossing criterion affects the theory at the level of non-universal properties such as the wetting transition phase boundary. 
For example, if (23) is adopted for $\rho_{1,\times}$, then $b_w^{(2)}(1) =1.138$ solves (29). 

For $a<1$, \eqref{Vexpagen} predicts that a critical wetting transition is possible when the leading coefficient, $b^2$, which denotes the ``distance" to the critical wetting phase boundary, tends to zero, so $b_w (a<1) =0$. The structure of the full interface potential \eqref{intpot} implies that critical wetting does indeed take place within the double-parabola approximation to the model. The equilibrium wetting layer thickness $\hat L$ diverges logarithmically in the manner
\begin{eqnarray}
\hat L \sim \frac{2a}{1-a}\left[ \ln\frac{1}{b} + \ln \frac{1}{a} + const.\right],
\end{eqnarray}
for $b \downarrow 0$. The spreading coefficient is predicted to vary as
\begin{eqnarray}
-S \propto b^{2/(1-a)},
\end{eqnarray}
implying
\begin{eqnarray}
2-\alpha_s = \frac{1}{1-a},
\end{eqnarray}
in full agreement with what was already analytically conjectured and numerically verified in \cite{Long}. 
Interestingly, as was also anticipated in \cite{Long}, there is no cross-over to a universal second-order wetting transition at $a=1/2$. Indeed, this is  conspicuous and emerges naturally in our new derivation, because the coefficient of $e^{-2 L}$ varies as $b^2$, so that the third term of $V(\hat L)$ near wetting behaves as $b^{2(1+a)/(1-a)}$, which is always sub-dominant with respect to the second term, no matter how small $a(>0)$ is taken. Recall that the second as well as the first term near wetting behave as $b^{2/(1-a)}$. This persistent non-universality is confirmed by considering the full $V(L)$ given in \eqref{intpot} and is in excellent agreement with the prediction of high-precision computations in \cite{Long} that critical wetting remains non-universal from $a \lesssim 1$ down to $a= 1/5$ at least.  Intriguingly, one of the implications of this theory is that critical wetting is of order {\em less than} 2 for $a < 1/2$ and asymptotically becomes first order in the limit $a \downarrow 0$, an exotic scenario already sketched in \cite{Long}.

For the infinite-order wetting transition, emerging in the limit $a \uparrow 1$, we retrieve precisely the predictions of \cite{Long},
\begin{eqnarray}\label{expsingula}
\hat L \propto \frac{1}{1-a},
\end{eqnarray}
implying $\beta_s = -1$, and 
\begin{eqnarray}\label{Ssingula}
-S \propto \left[\frac{b}{b^{(2)}_w(1)}\right]^{2a/(1-a)} \propto e^{-C/(1-a)}, \;\;\mbox{with} \;C>0,
\end{eqnarray}
confirming the infinite-order character of the transition. All these results are essentially independent of the numerical value given to $\rho_{1,\times} \in (0,1)$, corroborating the robustness of the crossing criterion we adopted to derive an interface potential.

\section{Thermal fluctuation effects on infinite-order wetting: linear functional RG approach}

Within local Hamiltonian theory the RG approach starts by considering an effective interface Hamiltonian of the form
\begin{eqnarray}\label{intham}
\frac{{\cal H}[\ell]}{kT} = \int dx \int dy \left \{ \frac{K_e}{2} (\nabla \ell (x,y))^2 + V(\ell(x,y)) \right \},
\end{eqnarray}
where, since the wetting layer is bounded by two similar interfaces whose thermal fluctuations are independent and additive, the (dimensionless) {\em effective surface tension} $K_e$ satisfies the following combining rule \cite{KardarInd,CM}
\begin{eqnarray}\label{efftension}
\frac{1}{K_e} = \frac{kT}{\xi^2}\left (\frac{1}{\sigma_{\alpha\beta}} + \frac{1}{\sigma_{\beta\gamma}} \right ) \equiv \frac{kT}{\xi^2}\frac{2}{\sigma},
\end{eqnarray}
with $\xi$ the bulk correlation length in the $\beta$ phase, and $\sigma = \sigma_{\alpha\beta} = \sigma_{\beta\gamma}$ (owing to the symmetry of the model).
Note that our definitions of ${\cal H}$ and $V(\ell)$ entail the following definition for the wetting parameter $\omega$ \cite{FH}
\begin{equation}
\omega = \frac{1}{4\pi K_e} = \frac{kT}{2\pi\sigma \xi^2},
\end{equation}
which is {\em twice} the value \eqref{omega} for a single interface unbinding from a flat wall. Therefore, we are dealing with significantly enhanced thermal fluctuations as compared to the usual SRCW problem in three dimensions!

We proceed in two stages. In the first stage, we keep only the leading terms in the large-$L$ expansion of $V(L)$, with $L \equiv A^{1/2} \ell$, and discuss the resulting singular behaviour at critical wetting. In the second stage, we include - as is physically required \cite{FH} - a term which mimics a soft repulsion penalizing the two interfaces  when attempting to cross each other (penalizing negative values of $L$). It would be reasonable physically to consider a hard repulsion, but the linear RG approach cannot properly handle an infinite potential. In the linear functional RG approach the renormalized interface potential is obtained by integrating out capillary wave fluctuations. This amounts to a convolution of the bare potential with a Gaussian of width $\delta$, where $\delta$ is the roughness of the fluctuating interface \cite{BHL,FH}. The width $\delta$, often referred to as the perpendicular correlation length of the interface, $\xi_{\perp}$, can be calculated using capillary wave theory \cite{CW1,CW2}. It is related to the parallel correlation length of the interface through
\begin{equation}\label{CW}
\xi_{\perp}   = \sqrt{ 2\omega \ln \xi_{\parallel}  },
\end{equation}
where all lengths are scaled with the bulk correlation length $\xi$. After these general considerations we now focus first on the regime of infinite-order transitions and therefore assume $0<b<b_w^{(2)}(1)$ and $a \lesssim 1$.

{\bf Stage 1}. To alleviate the notation we model the bare potential as
\begin{align}\label{simple}
V(L) =  \left \{  \begin{array}{l} 
  - D e^{- L} + B e^{- L/a}, \; \mbox{for} \; L > 0\\
 0, \; \mbox{for} \; L < 0,
\end{array}
\right . 
\end{align}
with $D$ and $B$ positive constants near infinite-order wetting, which satisfy $B/D \approx (2 \rho_{1,\times}/b)^2 = (b_w^{(2)}(1)/b)^2 >1$.

At this stage no care is taken to exclude those (rare) capillary wave fluctuations that would lead to crossings of the two wandering interfaces. Since the calculations closely follow those outlined in \cite{FH} we do not repeat them here, but limit ourselves to reporting the results and mentioning only those calculational details that merit special attention.
The renormalized potential  reads
\begin{equation}\label{renormalized}
V_R(L) = - D \xi_{\parallel}^{\omega} e^{- L} + B \xi_{\parallel}^{a^{-2}\omega} e^{-  L/a}, \; \mbox{for large} \; L,
\end{equation}
which has the same form as $V$ but with multiplicatively renormalized coefficients. We recall that all lengths are implicitly scaled with the bulk correlation length $\xi$. This form of renormalized potential is valid under the following conditions: $L > 2 \omega \ln \xi_{\parallel}$ for the first term and $L > (2/a) \omega \ln \xi_{\parallel}$ for the second term. Since $a<1$ the latter condition implies the former, but note that for $a \uparrow 1$ the conditions become coincident. This condition expresses that the two interfaces fluctuate far enough from each other to avoid mutual collisions and defines the so-called weak fluctuation regime. We will see shortly that this corresponds to a definite range of $\omega$.

Minimization of $V_R(L)$ leads to the {\em renormalized} equilibrium wetting layer thickness, which we denote by $\hat L_{R}$, and the second derivative of $V_R(L)$ evaluated in $\hat L_{R}$ provides the parallel correlation length $\xi_{\parallel}$ through $\ddot V_R(\hat L_R) \propto \xi_{\parallel}^{-2}$, as outlined in \cite{FH}. Combining the resulting relations and eliminating the dependence on the ratio $B/D$, which is merely a constant $(>1)$ upon approach of the infinite-order wetting transition, leads to the following relation between  $\hat L_{R}$ and $\ln \xi_{\parallel} $,
\begin{equation}\label{elleqlnxi}
\hat L_{R} = \ln\left(\frac{1}{a}-1\right)  + (2+ \omega)  \ln \xi_{\parallel} + const., 
\end{equation}
where we show only the terms that matter in the limits $a \uparrow 1$ and $\xi_{\parallel} \rightarrow \infty$.
A similar calculation leads to the following relation between $\hat L_{R}$ and $1-a$, which we present in the form, valid for $a \lesssim 1$,
\begin{equation}
e^{-\hat L_{R}}  \approx  (B/D)^{-(2+\omega)/[(2-\omega)(\frac{1}{a}-1)]}  [const.(\frac{1}{a}-1)]^{2\omega/(2-\omega)},
\end{equation}
which implies, asymptotically for $a \uparrow 1$,
\begin{equation}
\hat L_R  \sim \frac{2+\omega}{(2-\omega)\left(\frac{1}{a}-1\right)} \ln \frac{B}{D} - \frac{2\omega}{2-\omega} \ln \left(\frac{1}{a}-1\right).
\end{equation}
Note that the first term determines the leading algebraic divergence of $\hat L_R$, with a critical exponent $\beta_s = -1$ that is unchanged with respect to the mean-field result. Also note that the amplitudes of both leading and subleading terms diverge for $\omega \uparrow 2$ (strong thermal fluctuations). 

We also derive the following relation between $\xi_{\parallel} $ and $1-a$, likewise valid  for $a \lesssim 1$,
\begin{equation}\label{xiparr}
\xi_{\parallel}   \approx  \left(\frac{B}{D}\right)^{1/[(2-\omega)(\frac{1}{a}-1)]}  \left[const.\left(\frac{1}{a}-1\right)\right]^{-1/(2-\omega)}, \,\,\mbox{with} \,\, B/D >1.
\end{equation}
Note that the first factor captures the leading exponential divergence, for $a \uparrow 1$, and the second factor embodies an algebraic divergence of the amplitude of this singularity, for fixed $\omega < 2$. 

The relation between $\xi_{\perp}$, defined through \eqref{CW}, and $1-a$ merits our special attention. Indeed, at the infinite-order wetting transition $\xi_{\perp}$ displays a universal algebraic divergence of the form
 \begin{equation}\label{xiperp}
\xi_{\perp}   \propto \sqrt{\frac{2\omega}{2-\omega}} \;\;(1-a)^{-1/2},
\end{equation}
implying the following result for the critical exponent of the thermally fluctuating interface width
 \begin{equation}\label{xiperp}
\nu_{\perp}  = 1/2.
\end{equation}
This is interesting because, to our knowledge, for other SRCW transitions in $d=3$, we invariably have $\nu_{\perp} = 0$(log). Note, once again, that the amplitude diverges for sufficiently strong fluctuations, i.e., for  $\omega \uparrow 2$.

Note that $\xi_{\parallel} $ displays an exponential singularity, while $\hat L_R$ and $\ln \xi_{\parallel}$ show an algebraic divergence, in the limit $a \uparrow 1$. Therefore, taking twice the logarithm of \eqref{xiparr} we can rewrite \eqref{elleqlnxi} in the more systematic form of an expansion in large $\xi_{\parallel} $,
\begin{equation}\label{elleqlnxisys}
\hat L_R =  (2+ \omega)  \ln \xi_{\parallel} - \ln \ln  \xi_{\parallel}  + ...
\end{equation}
This allows us to check the self-consistency requirement, $\hat L_R > (2/a) \omega \ln \xi_{\parallel}$, for $\hat L_R, \ln \xi_{\parallel}  \rightarrow \infty$,
\begin{equation}\label{selfcon}
\hat L_R /  \ln \xi_{\parallel} \approx  2+ \omega > (2/a)  \omega, 
\end{equation}
which leads to the following condition on $\omega$,

\begin{equation}\label{selfcon}
\omega < \frac{2}{(2/a)-1}  \approx 2.
\end{equation}

We conclude that there is only {\em one} ``weak" (and no ``intermediate") fluctuation regime for infinite-order wetting, defined by $0 < \omega < 2$. Note that for $\omega \uparrow  2$ an {\em additional} exponential singularity develops in $\xi_{\parallel} $, implying a wetting transition of doubly-infinite order. The ``strong" fluctuation regime, $\omega > 2$, that lies beyond this threshold appears interesting, too, but falls outside the scope of our paper.

We proceed to examine the leading singularity of the spreading coefficient at infinite-order wetting in the presence of thermal fluctuations and to check hyperscaling. Evaluating the interface potential at the equilibrium wetting layer thickness leads to
\begin{equation}\label{Ssing}
-S \approx \left(\frac{B}{D}\right)^{-2/[(2-\omega)(\frac{1}{a}-1)]}  \left[const.\left(\frac{1}{a}-1\right)\right]^{2/(2-\omega)}, \,\,\mbox{with}\,\, B/D \approx [b_w^{(2)}(1)/b]^2 >1.
\end{equation}

We observe that $-S \propto  \xi_{\parallel}^{-2}$, which implies that hyperscaling holds. (We recall that hyperscaling amounts to the exponent equality $2 - \alpha_s = (d-1)\nu_{\parallel}$.) It can readily be seen that hyperscaling holds quite generally for the interface potentials of the type that we study, since the second derivative of $V_R$ is proportional to $V_R$ itself, when both are evaluated at $\hat L_R$. 

We conclude that the main characteristics (the critical exponents) of the mean-field infinite-order wetting transition ($2-\alpha_s = \infty$ and $\beta_s = -1$) are {\em robust} to thermal fluctuations, i.e., independent of the value of $\omega$,  provided $0 < \omega < 2$. In addition, we remark that the {\em power} of $1-a$ in the argument of the exponential singularity (i.e., in the exponent of $B/D$ in \eqref{Ssing}) is also robust to thermal fluctuations and preserves its mean-field value $-1$, as in \eqref{Ssingula}, whereas the {\em amplitude} of $1/(1-a)$ in this argument does depend on $\omega$.  

{\bf Stage 2}. We now progress towards a physically better founded model and add a non-zero soft repulsion between the two fluctuating interfaces by augmenting the bare interface potential with a positive constant $E$ at $L \leq 0$; 
\begin{align}\label{simpleE}
V(L) =  \left \{  \begin{array}{l} 
 - D e^{- L} + B e^{- L/a}, \; \mbox{for} \; L > 0 \\ 
             E, \; \mbox{for} \; L<0,
\end{array}
\right . 
\end{align}
Since the constant $E$ renormalizes (approximately) to a Gaussian, the renormalized potential now reads, for $\omega < 2$,
\begin{equation}\label{renormalizedC}
V_R(L) = - D \xi_{\parallel}^{\omega} e^{- L} + B \xi_{\parallel}^{a^{-2}\omega} e^{-  L/a} + \frac{E}{L}\sqrt{\frac{\omega\ln \xi_{\parallel}}{\pi} } \, e^{ -L^2 /(4 \omega \ln \xi_{\parallel})}, \; \mbox{for large} \; L
\end{equation}
We must now check whether the solution for $\hat L_R$ obtained at stage 1 is still valid. This is the case provided the Gaussian remains small compared to the other two terms, when the solution found at stage 1, being \eqref{elleqlnxisys}, is inserted in $V_R(L)$. One verifies that the first term scales as $\ln \xi_{\parallel}/ \xi_{\parallel}^2$, the second as $(\ln \xi_{\parallel})^{1/a}/ \xi_{\parallel}^2$, with $a \approx 1$ close to the wetting transition, and the third term scales as $(\ln \xi_{\parallel})^{(2+\omega)/2\omega}/\xi_{\parallel}^{(2+\omega)^2/4\omega}$. Since $(2+\omega)^2 \geq 8 \omega$ for all $\omega$ (equality for $\omega =2$), the third term is negligible compared to the other two for $0 < \omega < 2$. In conclusion, the properties of the (renormalized) infinite-order wetting transition are not sensitive to whether or not a soft repulsion is added to the interface potential. Calculational stages 1 and 2 are equivalent for this particular wetting transition.

\section{Thermal fluctuation effects on non-universal wetting ($a < 1$): linear functional RG approach}

In this Section we renormalize the wetting transition of continuously varying {\em finite} order. We thus concentrate on the parameter ranges $0<a<1$ and $b \ll 1$. Recall that now $b_w = 0$. In doing so we will reproduce the results of Hauge and Olaussen (HO), who pioneered RG corrections to a wetting transition that has non-universal character at mean-field level \cite{HO}. They studied the weak fluctuation regime. Parry {\em et al.} \cite{bern} also examined the intermediate fluctuation regime for that transition and found that the universality with respect to asymmetry is restored, while the non-universality with respect to the wetting parameter $\omega$ persists. Why do we revisit these non-universalities, given that they have been explored before? The interface potential we encounter in our theory is qualitatively different from those of previous works, in that the coefficients of powers of $\exp(-L)$ all vanish at wetting, while the coefficients of powers of $\exp(-L/a)$ do not. This has important consequences for the size of the non-universal regime in the wetting phase diagram. Further, we pay special attention to the fact that in our case of two fluctuating interfaces thermal fluctuation effects are significantly enhanced, since - as we outlined in the previous section - the value of $\omega$ is roughly doubled relative to that for wetting at a planar undeformable wall. As in the previous section, we proceed in two calculational stages or ``models". At stage 1 we consider only the  leading terms, for large $L$, in the interface potential, and at stage 2 we add a soft repulsion to discourage interface wandering from visiting the unphysical domain $L < 0$.

{\bf Stage 1}.
We start from the bare interface potential based on \eqref{Vexpagen}
\begin{align}\label{simple}
V(L) =  \left \{  \begin{array}{l} 
  - D e^{- L} + B e^{- L/a} + De^{- 2L} , \; \mbox{for} \; L > 0\\
 0, \; \mbox{for} \; L < 0,
\end{array}
\right . 
\end{align}
with $D \propto b^2$ the ``distance" to the critical wetting transition at $b_w=0$, and $B \propto \rho_{1,\times}^2/a$ a positive constant. 

Its renormalized counterpart is given by
\begin{equation}\label{renormalizednonu}
V_R(L) = - D \xi_{\parallel}^{\omega} e^{- L} + B \xi_{\parallel}^{a^{-2}\omega} e^{-  L/a} + D \xi_{\parallel}^{4\omega} e^{-2 L}, \; \mbox{for large} \; L,
\end{equation}
and is valid in the weak fluctuation regime $0 < \omega < \omega_1$, where $\omega_1$ is to be calculated. This form of renormalized potential is valid under the following conditions: $L > 2 \omega \ln \xi_{\parallel}$ for the first term, $L > (2/a) \omega \ln \xi_{\parallel}$ for the second term and $L > 4 \omega \ln \xi_{\parallel}$ for the third term. Note that, for $a$ below some value, the next-to-leading term in $V(L)$ {\em may} ``cross" some higher-order term and the transition may {\em lock in} to one of the universal critical wetting kind. At mean-field level, however, this does not happen for our model (see Section III). In order to exclude any possible complications of this sort, we limit ourselves for the time being to the range $a_c < a < 1$, where $a_c$ is to be calculated.  It then suffices to work with the {\em first two terms} in \eqref{renormalizednonu} to derive the singularities at wetting. 

Minimization of the first two terms of $V_R(L)$ leads to the following relation between the (diverging) equilibrium wetting layer thickness and the (diverging) quantities $B/D$ and $\ln \xi_{\parallel}$,

\begin{equation}\label{elleqlnxinon}
\hat L_R = \frac{a}{1-a} \left ( \ln \frac{B}{aD} + (\frac{1}{a^2} -1) \omega \ln \xi_{\parallel} \right ),
\end{equation}
and, combined with the evaluation of the second derivative, or ``curvature", in the minimum of $V_R$, this provides an independent  relation between the parallel correlation length and, essentially, the ``field" $1/D$ by which wetting can be induced in the limit $1/D \rightarrow \infty$,
\begin{equation}\label{secondder}
\ln \xi_{\parallel}   =  \frac{1}{2-\omega/a}  \ln  \frac{(B/aD)^{a/(1-a)}}{(1-a)D/a} + const.,
\end{equation}

These two auxiliary equations, \eqref{elleqlnxinon} and \eqref{secondder}, now provide the asymptotic relation between $\hat L_R$ and $\ln \xi_{\parallel} $ close to wetting,
\begin{equation}\label{elleqlnxinons}
\frac{\hat L_R}{\ln \xi_{\parallel}} \approx 2a + \omega/a,
\end{equation}
The sufficient condition for the validity of the first two terms in \eqref{renormalizednonu}, being $L > (2/a) \omega \ln \xi_{\parallel}$, requires this ratio \eqref{elleqlnxinons} to exceed $2  \omega/a$, and signifies that the interfaces fluctuate sufficiently far from one another to avoid collisions. It defines the {\em weak fluctuation regime},
\begin{equation}\label{selfconnon}
\omega < 2 a^2 \equiv \omega_1
\end{equation}
As expected, we retrieve the same $\omega_1$ as in the model studied by HO, which further reduces, for $a=1/2$, to the famous value 1/2 found in \cite{FH}. 

We proceed to study the critical exponent $\nu_{\parallel}$ of the correlation length $\xi_{\parallel}$ parallel to the unbinding interfaces.
Defining, as is standardly done,
\begin{equation}\label{xiexp}
\xi_{\parallel} \propto D^{-\nu_{\parallel}},
\end{equation}
and using \eqref{secondder}
we find
\begin{equation}\label{nuparallelweak}
\nu_{\parallel} = \frac{1}{2(1-a)}\frac{1}{1-\omega/(2a)},
\end{equation}
as in the model of HO \cite{HO}, and consequently we obtain $\nu_{\perp} = 0($log), in view of \eqref{CW}.
The first factor in the r.h.s. of \eqref{nuparallelweak} defines the mean-field contribution (the limit $\omega = 0$) and last factor contains the thermal fluctuation correction. Note that the order of the transition increases from its mean-field value when $\omega$ is increased, i.e., when  thermal fluctuations gain importance. One easily verifies, along the lines of the derivation given in \cite{FH}, that hyperscaling holds so that $2-\alpha_s = 2 \nu_{\parallel}$.  

We now show that in the regime of validity of the first three terms in \eqref{elleqlnxinons}, defined by the sufficient condition $L > (2/a) \omega \ln \xi_{\parallel}$ {\em and} $L > 4 \omega \ln \xi_{\parallel}$, i.e., $\omega < 2 a^2$ {\em and} $(4a-1) \omega < 2 a^2$, the third term evaluated in $\hat L_R$ is negligible compared to the first two, not only for $1/2 < a < 1$ but for all $0<a<1$,  so $a_c=0$. Indeed, using \eqref{elleqlnxinons} and \eqref{xiexp} to examine how the various terms in $V_R(\hat L_R)$ scale near wetting, we obtain that the first two terms scale as $\xi_{\parallel}^{-2}$ while the third term scales as $\xi_{\parallel}^{-2-2a+(3-1/a)\omega}$, which is subdominant provided $(3a-1)\omega < 2a^2$. This is indeed fulfilled, since the former condition $(4a-1) \omega < 2 a^2$ implies this one for all $a>0$. 

The question now remains whether $a_c=0$ also holds for $L > (2/a) \omega \ln \xi_{\parallel}$ but $L < 4 \omega \ln \xi_{\parallel}$, so that the third term in \eqref{renormalizednonu} must be replaced by a suitably renormalized version of the bare term $De^{- 2L}$. This question is obviously only meaningful for $a>1/2$, since for $a<1/2$ the validity of the second term implies that of the third and we are back to the previous conclusion reached in case all three terms in \eqref{renormalizednonu} are valid. The suitably renormalized interface potential now reads, the third term being a Gaussian (with a positive amplitude) \cite{FH},
\begin{equation}\label{renormalizedGauss}
V_R(L) = - D \xi_{\parallel}^{\omega} e^{- L} + B \xi_{\parallel}^{a^{-2}\omega} e^{-  L/a}+ \frac{D}{\sqrt{4\pi\omega \ln \xi_{\parallel}} }\frac{1}{2-L/ (2\omega  \ln \xi_{\parallel}) } e^{- L^2 / (4 \omega \ln \xi_{\parallel}) } + ...,
\end{equation}
with, as before, $D \propto b^{2}$. We now obtain that the first two terms scale as $\xi_{\parallel}^{-2}$ while the third term scales as $\xi_{\parallel}^{-(2+[(2a-1)\omega/(2a) - a]^2/\omega)} $, which is  subdominant because the exponent is less than -2. We conclude $a_c=0$. In the model of Stage 1, therefore, the critical wetting transition for $a<1$ is (doubly) non-universal down to $a=0$ in the weak-fluctuation regime.

We now turn to the intermediate fluctuation regime, $\omega_1 <\omega < \omega_2$, where $\omega_2$ is still to be determined and $\omega_1$ is to be checked on its self-consistency. In this regime $L > 2 \omega \ln \xi_{\parallel}$ and $L < (2/a) \omega \ln \xi_{\parallel}$. 
With these conditions the repulsive next-to-leading term of the interface potential is renormalized to a Gaussian (with a positive amplitude), and the renormalized $V(L)$ reads
\begin{equation}\label{renormalized2}
V_R(L) = - D \xi_{\parallel}^{\omega} e^{- L} +  \frac{B}{\sqrt{4\pi\omega \ln \xi_{\parallel}} }\frac{1}{(1/a)-L/ (2\omega  \ln \xi_{\parallel}) } e^{- L^2 / (4 \omega \ln \xi_{\parallel}) } + ...,
\end{equation}
with, as before, $D \propto b^{2}$ and $B$ a positive constant. 

Minimization of the interface potential and determination of the curvature (second derivative) is now straightforward. The calculations are simplified by legitimately ignoring the $L$-dependence of the amplitude (prefactor) of the second term, which amounts to dropping higher-order terms in $1/L$, as noted in \cite{FH}. After some algebra we obtain
\begin{equation}\label{one}
\hat L_R^2 = 8 \omega  (\ln \xi_{\parallel}) ^2 - 2 \omega \ln \xi_{\parallel} \ln \ln \xi_{\parallel} + ...,
\end{equation}
and
\begin{equation}\label{two}
\hat L_R = (2+  \omega)  \ln \xi_{\parallel}  + \ln D   + ...,
\end{equation}
where the dots stand for constants or terms that diverge more weakly than 
the ones that are shown. In line with the results of \cite{FH}, we assume the following formal relation between $\ln \xi_{\parallel}$ and $D$, in the limit $D \downarrow 0$,
\begin{equation}\label{formal}
\ln \xi_{\parallel} = f_1 \ln 1/D + f_2 \ln \ln 1/D + ...
\end{equation}
Applying this to leading order to \eqref{one} and \eqref{two} we get
\begin{equation}\label{nothing}
1/\nu_{\parallel} \equiv  1/f_1 = 2 + \omega  - \sqrt{8\omega} =  (\sqrt{2} - \sqrt{\omega})^2
\end{equation}
which reproduces the result  of \cite{FH}.
Note that, in contrast with \eqref{nuparallelweak}, in this intermediate fluctuation regime $\nu_{\parallel}$ is independent of the asymmetry variable $a$.
We now check the self-consistency condition on the range of the interface fluctuations, $L > 2 \omega \ln \xi_{\parallel}$ and $L < (2/a) \omega \ln \xi_{\parallel}$,  and obtain
\begin{equation}\label{selfconnonboth}
2a^{2}  < \omega < 2,
\end{equation}
confirming the previously found expression \eqref{selfconnon} for $\omega_1$ and the expected upper limit $\omega_2 = 2$ beyond which the strong fluctuation regime sets in, characterized by an essential singularity (the infinite-order transition induced by thermal fluctuations).

It is easy to see that these results are not affected by the presence of a third term in \eqref{renormalized2}. Assume first $L > 2 \omega \ln \xi_{\parallel}$ and $L < (2/a) \omega \ln \xi_{\parallel}$, but $L < 4 \omega \ln \xi_{\parallel}$, in which case the third term is a Gaussian of the same form as the (Gaussian) second term, but with an amplitude $D$ that vanishes at wetting. The third term is then negligible compared to the second. In the opposite case, $L > 2 \omega \ln \xi_{\parallel}$ and $L < (2/a) \omega \ln \xi_{\parallel}$, but $L > 4 \omega \ln \xi_{\parallel}$, possible for $a<1/2$, and $2a^2 < \omega < 1/2$, the third term is exponentially decaying and takes the same form as the third term in \eqref{renormalizednonu}. We now obtain that the first two terms in the renormalized potential scale as $\xi_{\parallel}^{-2}$ while the third term scales as $\xi_{\parallel}^{-(2+\sqrt{8\omega} - 3 \omega) }$, which is  subdominant because the exponent is less than -2 provided $\omega < 8/9$, which is guaranteed by $\omega < 1/2$.

We conclude that in the intermediate fluctuation regime the mean-field non-universality is washed out or overruled by thermal fluctuation effects, so that only the non-universality induced by thermal fluctuations remains. Note that the upper limit of the intermediate fluctuation regime is $\omega = 2$, for which $\nu_{\parallel}$ diverges.

In closing this stage of the calculations we comment on the behaviour of the wetting layer thickness at wetting. At non-universal critical wetting $\hat L_R$ diverges logarithmically for  $1/D\rightarrow \infty$.  In particular, in the weak fluctuation regime we obtain, combining \eqref{secondder} and \eqref{elleqlnxinons},
\begin{equation}\label{leqqweak}
\hat L_R = \frac{2 + \omega /a^{2}}{(1/a-1)(2-\omega/a)} \ln \frac{1}{D}+ ...
\end{equation}
which for  $a=1/2$ reduces to the result of \cite{FH}.

Likewise, in the intermediate fluctuation regime we combine \eqref{one} and \eqref{two} and obtain, first, the dependence of $\hat L_R$ on $\ln \xi_{\parallel}$, from which the parameter $a$ has disappeared as expected (partial restoration of universality),
\begin{equation}
\hat L_R =  \sqrt{8\omega}   \ln \xi_{\parallel}             - \frac {\omega \ln \ln \xi_{\parallel} }{\sqrt{8\omega}} + ...,
\end{equation}

Working out \eqref{formal} this leads to
\begin{equation}
\hat L_R = \frac{  \sqrt{8\omega} }{ 2 + \omega -  \sqrt{8\omega}}  \ln \frac{1}{D} - \frac{(2+\omega)  \omega}{  \sqrt{8\omega}(2 + \omega -  \sqrt{8\omega})}\ln \ln \frac{1}{D} + ... ,
\end{equation}
These relations reproduce the results of \cite{FH}. 

{\bf Stage 2}.
We now start from the bare interface potential augmented with the soft repulsion, \eqref{simpleE}, and consider its renormalized counterpart \eqref{renormalizedC} valid in the weak fluctuation regime $0 < \omega < \omega_1$, where $\omega_1$ is, again, to be calculated. As in stage 1 we focus on the asymmetry range susceptible of critical wetting, $0<a<1$. As noted before, this form of the renormalized interface potential \eqref{renormalizedC} is valid, provided $L > (2/a) \ln \xi_{\parallel}$, which is a {\em necessary} condition. However, it is possible that the solution obtained at stage 1 is only valid in a part of the range of $\omega$ defined by this  condition, because the third term in \eqref{renormalizedC} may become more important than the second. To examine this, we recall the solution obtained at stage 1, as given in \eqref{elleqlnxinons}. When we insert this solution into the various terms of \eqref{renormalizedC}, taking into account the explicit form of the critical exponent \eqref{nuparallelweak} of the parallel correlation length, we find that the first two terms scale as $\xi_{\parallel}^{-2}$, while the third term is negligible compared to the first two, {\em provided}
\begin{equation}
(\hat L_R / \ln \xi_{\parallel} )^2 >   8\omega 
\end{equation}
Calling $\omega_{c1}$ the  value of $\omega$ that solves this inequality as an equality, we obtain
\begin{equation}
\omega_{c1} = 2  a^2,  
\end{equation}
identical to the $\omega_1$ found in stage 1. We  conclude that the  weak fluctuation regime is, again, defined by $0 < \omega <  2a^2$. 

For $\omega > 2  a^2$ the Gaussian repulsion dominates the exponential one (second term in \eqref{renormalizedC}) and the correct solution can be obtained by retaining only the first, attractive, term and the third. This renormalized interface potential, valid in the intermediate fluctuation regime reads
\begin{equation}\label{renormalizedCintermediate}
V_R(L) = - D \xi_{\parallel}^{\omega} e^{- L} +  \frac{E}{L}\sqrt{\frac{\omega\ln \xi_{\parallel}}{\pi} } \, e^{ -L^2 /(4 \omega \ln \xi_{\parallel})},
\end{equation}
and is independent of the parameter $a$ and therefore independent of the asymmetry variable in our model. The validity of this description is limited to $L / \ln \xi_{\parallel} > 2 \omega$, as noted previously. In this intermediate fluctuation regime the solution is found to be akin to that obtained in \cite{FH} and is given by 
\begin{equation}\label{leqqinterC}
\hat L_R    =   \sqrt{ 8\omega}(\ln \xi_{\parallel}- \frac{1}{8} \ln \ln \xi_{\parallel} + ...), 
\end{equation}
and
\begin{equation}
1/\nu_{\parallel} = 2  + \omega  - \sqrt{8\omega}
\end{equation}
It is straightforward to check that this expression is continuous across $\omega_{c1}$. At $\omega_{c1}$ it coincides with that found in the weak fluctuation regime. 

In the intermediate fluctuation regime we thus retrieve the ``old" results, which are universal with respect to the asymmetry, and non-universal in the sense that they depend on the thermal fluctuation strength as measured by $\omega$. This intermediate fluctuation regime is applicable in the range $\omega_{c1} < \omega < \omega_{c2} = 2$. For $\omega > 2$ the first term in \eqref{renormalizedC} also becomes a Gaussian and we enter the strong fluctuation regime already discussed in \cite{FH}. 

As far as the wetting layer thicknesses are concerned, we observe that the stage-2 calculations reproduce the results \eqref{elleqlnxinons} and  \eqref{leqqweak} obtained at stage 1 in the weak fluctuation regime, as well as \eqref{leqqinterC} and the ensuing dependence of $\hat L_R$ on $D$ obtained in \cite{FH} in the intermediate fluctuation regime. 

\section{Conclusions}
We have studied, within a local interface Hamiltonian theory, the effect of thermal fluctuations on  wetting phase transitions of infinite order and of finite, continuously varying, order. At mean-field level, these transitions were uncovered in a density-functional model for a system with short-range forces and a two-component order parameter. The main new results are the following. Using linear functional renormalization group (RG) calculations we have shown that the infinite-order transitions are robust with respect to the inclusion of thermal fluctuation effects. The exponential singularity of the surface free energy at wetting, characterized by the critical exponent $2-\alpha_s = \infty$ and the  algebraic divergence of the wetting layer thickness, characterized by the critical exponent $\beta_s = -1$, are not modified provided $\omega < 2$, with $\omega$ the dimensionless wetting parameter that measures the strength of thermal fluctuations. The interface width, or perpendicular correlation length, $\xi_{\perp} $ diverges algebraically and universally, characterized by the critical exponent $\nu_{\perp} = 1/2$. Under strong fluctuations, for $\omega \uparrow 2$, the order of the wetting transition becomes doubly infinite and a new regime is entered, which has not been studied here.

As regards the non-universal critical wetting transitions of finite but continuously varying order, we recall that at mean-field level the critical exponent of the surface free energy singularity at wetting depends on the asymmetry parameter $a$ of the model. We also recall that this dependence persists through the entire range $0<a<1$ available to this parameter. Unlike in other, but similar, mean-field models, the transition does not lock in to a universal second-order wetting transition at some value of $a$. We have found, using linear functional renormalization group (RG) calculations, that the mean-field non-universality persists in the weak fluctuation regime $0<\omega < 2a^2$ and that a second non-universality with respect to thermal fluctuations adds on to this. In contrast, in the intermediate fluctuation regime $2a^2<\omega < 2$, universality with respect to the asymmetry parameter $a$ is restored and $\alpha_s$  depends on $\omega$ alone. For $\omega \uparrow 2$ we enter the strong fluctuation regime where the wetting transition is predicted to be of infinite order. For the wetting transitions of finite and continuously varying order, our calculations have reproduced various known results \cite{FH,HO,bern}. 

In a follow-up work we envisage to meet the challenge of deriving the interface potential and studying fluctuation effects on the singularities at wetting in this model, from a fully nonlocal interface Hamiltonian theory. This will allow us to capture better the effects of two thermally wandering interfaces, typical for fluid systems without rigid flat walls.

\section{Acknowledgements}
J.O.I. thanks Joachim Krug for his interest in these results. H.H. is Research Assistant of the Fund for Scientific Research of Flanders (FWO-Vlaanderen). J.O.I. and H.H. are supported by KU Leuven Research Grant OT/11/063.

\bibliography{apssamp}

\begin{thebibliography}{99}
\bibitem{note} More precisely, the total (scaled) wetting layer thickness, which we now denote by $L$, amounts, in \cite{Long}, to $2L + 2Z_{\times}$, where $Z_{\times}$ is a ``crossing distance" that remains finite for $L \rightarrow \infty$ in \cite{Long}.
\end{thebibliography}

\end{document}